\documentclass[12pt]{amsart}
\usepackage{geometry, graphicx} 
\title{Umbral dots observed in photometric images taken with 1.6 m solar telescope}
\author{Aleksandra Andic$^{1,2}$}
\date{November 7th, 2011} 

\begin{document}

\maketitle
\tableofcontents

\section{Abstract}
Umbral dots (UDs) were observed using the 1.6 meter solar telescope. Achieved conventional diffraction limit in the TiO 705.68 nm spectral line used was $0.1''$. The 418 UDs were analysed. Median diameter was $0.5''$ and median intensity difference between darkest part of the UD's background and brightest part of the UDs was 37\%. Despite to achieved resolution, no UDs substructures were visible. The analysed UDs appeared to be circular. \par
\vskip45mm

\small{$^1$Department of Astronomy, New Mexico State University\\ P.O.Box 30001, MSC 4500, Las Cruces, NM 88003, USA}\par
\small{$^2$Big Bear Solar Observatory\\ 40386 N. Shore. Ln., Big Bear City, CA 92314, USA}

\clearpage

\section{Introduction}

 Umbral dots (UDs), are small-scale structures distributed across the umbra. Modelling showed that appearance of the UDs is natural consequence of the magnetoconvection under the conditions set by a strong magnetic field (Sch\"{u}ssler and V\"{o}gler, 2006).  This was confirmed by Rempel et al. (2009), Cheung et al. (2010) and Bharti et al. (2010), however all authors used MURaM code (V\"{o}gler et al. 2005), same as Sch\"{u}ssler and V\"{o}gler (2006). Heinemann et al. (2007) state that UDs are caused by overturning convection using PENCIL code (Heinemann et al. 2006). Heinemann et al. (2007) also saw indications of the umbral dots substructure, similar to those showed in Schu\"{u}ssler and V\"{o}gler (2006). \par
  Sch\"{u}ssler and V\"{o}gler (2006) and consequently Bharti et al. (2010) observed in their simulations that resulting bright features have a horizontally elongated form with a central dark lane. Similar dark lane was also observed by Heinemann et al. (2007) in their simulations. Sch\"{u}ssler and V\"{o}gler (2006) found that simulated UDs correspond well to the observed UDs in the central parts of the umbrae, when brightens, lifetime and size are in question. Bharti et al. (2010) could not reproduce small, short-lived UDs. The UDs from their simulation existed for 25 to 28 minutes, covered averaged areas of 0.08 to 0.14 Mm$^2$, with brightness of  1.6 to 1.7 factor larger than surrounding dark background with the corresponding values for the continuum brightness at 630 nm of 2.6 to 2.9. \par
Rimmele (2008), Riethm\"{u}ller et al. (2008), and Ortiz et al. (2010) stated that they seen a signature of UDs dynamics as described by the Sch\"{u}ssler and V\"{o}gler (2006) model. Riethm\"{u}ller et al. (2008) said that although the vertical cuts of their data sets and inversion data agree with the model, they could not detect the strong down-flows associated with the central dark lane. They reasoned that such observational result is caused by the limited spatial resolution of their data of $\sim 0.32"$. Ortiz et al. (2010), with the resolution of 0.14" saw the substructure of the UD in the form of dark lanes, although not all UDs possessed the dark lane. The estimated size of the substructures matches the achieved resolution. At similar resolution Rimmele (2008)  saw some of the UDs with the dark substructure of the estimated size $\sim  0.12''$. Also, Rimmele (2008) observed that if the UD was located near the edge of the umbra the dark feature would extend out from the bright dot, with the appearance of a comet-like tail. \par
 Rimmele (2008) noted that UDs have lifetime close to 30 minutes, as predicted  by the Sch\"{u}ssler and V\"{o}gler (2006) model. Bharti et al. (2010) similarly predicted that the average UDs lifetime is between $25$ and $28$ minutes. On the other hand, Hamedivafa (2008) stated that the average lifetime of the UDs is between $7$ and $10$ min, while Watanabe et al. (2009) measured the average lifetime as $7.3$ minutes. \par 
This paper presents a preliminary analysis of the UDs observed with the $1.6$ meter telescope. Results presented here give  an additional insight in behaviour of the UDs, and can serve as a evaluation data for the observing the umbra with the various spectral lines and bands. 

\section{Data and analysis}

The data set used was obtained on 22 September 2010 using the photometry with the broad band filter centred at TiO 705.68 nm spectral line at New Solar Telescope (NST) at Big Bear Solar Observatory (Cao et al. 2010). The TiO broadband filter was chosen due to the technical reasons. Its use in optical path was suitable for ongoing engineering work. The work on the instrumental installation takes priority over obtaining of the scientific data  during commission phase. The line parameters of the molecular lines are suitable for umbral observation (Sinha \& Tripathi, 1991a, 1991b). The TiO line is molecular spectral line and hence should be suitable. The broad passband of the  filter used allowed us to observe the photosphere at the level close to $\tau_{500}=1$.   
\par
The target of the observation was active region NOAA AR 11108. The data sequence consists of 141 bursts with 100 images in each burst. The exposure time for an individual frame was $1$ ms. The  cadence of the data, between bursts, is $15$~s. The time series covers the time interval of $\sim 35$ minutes. The data were acquired in the morning with fairly constant average seeing levels. Low order adaptive optic was used. The images have a sampling of $0.034"$ per pixel, which resulted in data oversampling.\par
In a work such this, where one tries to observe  structures at the edge of the diffraction limit of the telescope,  decision which criterium will be used to describe  telescope resolution has to be made. Conventional diffraction resolution ($\lambda/D$) gives resolution of $0.09''$ for this spectral line, while at the same time Rayleigh criterion gives a value of $0.11"$, and Sparrow's criterium gives $0.086"$. The data were obtained using the instrumentation  located in a Coud\'{e} room utilising the newly installed low order adaptive optic (AO) system. With the dataset obtained utilising an AO system, together with the post image processing it is possible resolve clearly  Sparrow's limit for the given spectral line. Cases of such good resolving are already know from the previous works, for example work of Berger et al. (2004) were authors managed to reach the resolution of $0.1"$ in G-band using $1$ meter telescope. However, during observational run for this dataset seeing was average, causing the high levels of noise in the umbrae. In order to increase the reliability of the dataset all structures below $0.09"$ are filtered out, in order to remove the potential noise artefacts. This procedure in effect changed the resolution of our dataset. The Sparrow criterium is set on $0.09"$, conventional criterium at $\sim 0.1"$, while Rayleigh criterium gives: $\sim 0.12"$. \par
The data  were speckle reconstructed based on the speckle masking method of von der L\"{u}he (1993) using the code described in W\"{o}ger et al. (2008). The images were filtered in space removing all features smaller than $0.09"$. Since Rimmele (2008) estimate size of the umbral substructures was $\sim 1.2''$, filtering out the structures smaller than $0.09''$ should not remove the umbral substructures. However, the filtering reduced the contrast of the images for 10\%. \par
The cadence of our reconstructed data provided us with a Nyquist frequency of $67$~mHz. After speckle reduction, images were co-  aligned using a Fourier co-aligning routine, which uses cross-correlation techniques and mean squared deviation to provide sub-pixel co-alignment accuracy. However, the sub-pixel image shifting was not implemented in order to avoid interpolation errors. Instead, the procedure was iterated $6$ times to achieve the best possible co-alignment. \par

Identification of the UDs in field of view was performed with  modified method applied by S\'{a}nchez Almeida et al. (2004). Instead of playing series back and forth to detect the UDs we used NAVE method (Chae \& Sakurai, 2008) to track the plasma flow of the individual UDs. Also, the analysis of UDs displacement and UDs tracking was performed using the NAVE with following restrictions: only UDs that appeared for more than $0.75$ minutes and were at least $0.09''$ in diameter entered this analysis. The constraining criteria and methodology of the small structures selection was described in detail in  Andic et al. (2011). To accurately measure dimensions of the UDs the measuring was done  when they were the brightest. For each UDs the time when they had a maximum of their intensity was chosen. The intensity profile from that instance was used and utilising a method of  full-width at half maximum (FWHM) the diameter was measured. In case when the UD was not spherical shape, the FWHM from the light profile along the longest part of UD was measured.\par

\section{Results}

\begin{figure}
\includegraphics[scale=0.9]{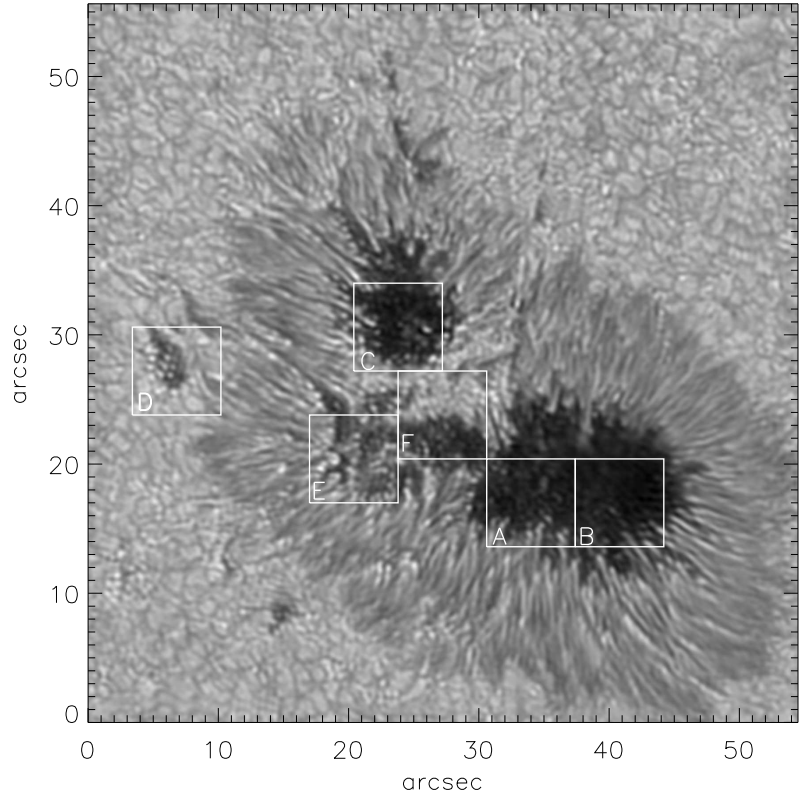} 
\caption{The observed NOAA AR 11108 with TiO 705.68 nm spectral line. Six white squares represent areas that are zoomed in Fig.\ref{pegzum}. 
}
\label{pega}
\end{figure}

An angular resolution of NST in TiO spectral line revealed plethora of the details in the umbra (Fig.\ref{pega}). 

\begin{figure}
\includegraphics[scale=0.8]{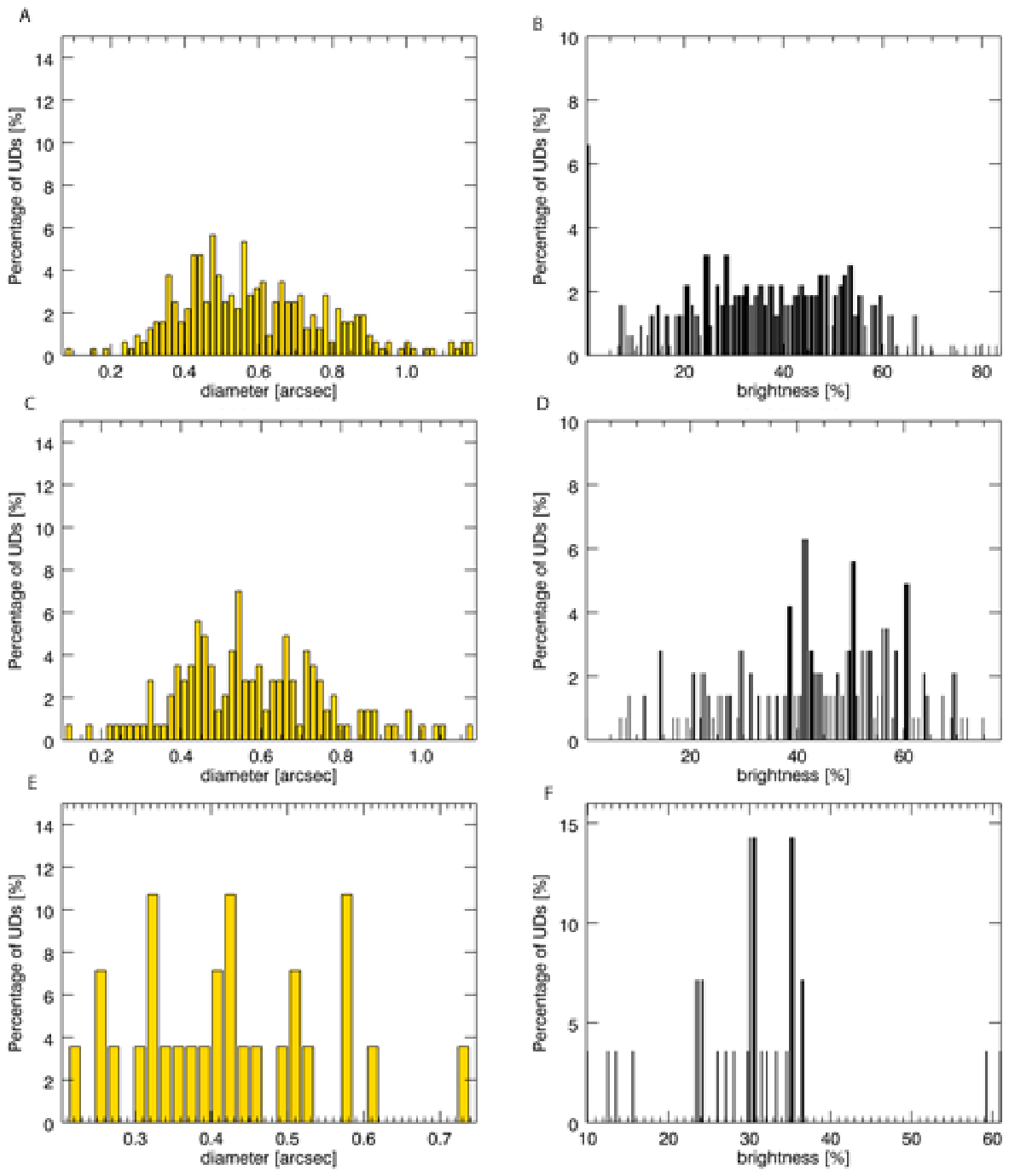} 
\caption{Results of the statistical analysis of the UDs. Panels A and B represent results for UDs from larger umbra, panels C and D for smaller umbra, while panels E and F represent the results for the UDs observed in pores. Panels A, C and E represent distribution of the diameters of the analysed UDs, while panels B,D and F distribution of the difference in the intensity between darkest pixel surrounding the UD and the brightest pixel in UD. }
\label{statistika}
\end{figure}

In this work, statistical analysis of 489 UDs is performed (Fig.\ref{statistika}). From that number, 318 were tracked in the bigger umbra, 143 in smaller and 28 in pores surrounding the spot.In all 3 regions the UDs had similar values for diameter and brightness, and persisted for a similar period of time. Median observable time for UDs from pores and smaller umbra was same, 35 minutes, while median observable time in larger umbra was shorter, 33.25 minutes. Diameter of the UDs varied from the median size in two umbras of $0.5''$ to median size of $0.4''$ for the UDs in pores. \par
Brightness was calculated as a percentage of  difference in the intensity between the brightest pixel in UD and the darkest pixel in the background surrounding the UD. The brightest UDs were observed in pores with the $43.9$\% difference in the intensity between darkest part of the background and brightest part of UD. The bigger umbra had the median value for the intensity difference (i.e. brightness) of $36.6$\% while in smaller umbra the median brightness was $30.6$\%. 

 Since the UDs substructure was observed with the instrumentation of the resolution  coarser than NST's one (Rimmele, 2008; Riethm\"{u}ller et al., 2008; Ortiz et al., 2010), there was an expectation to see clearly UDs substructures in this dataset. However, not a single UDs in this dataset had a dark lane as described in  model of Sch\"{u}ssler and V\"{o}gler (2006). \par

\begin{figure}
\includegraphics[scale=0.9]{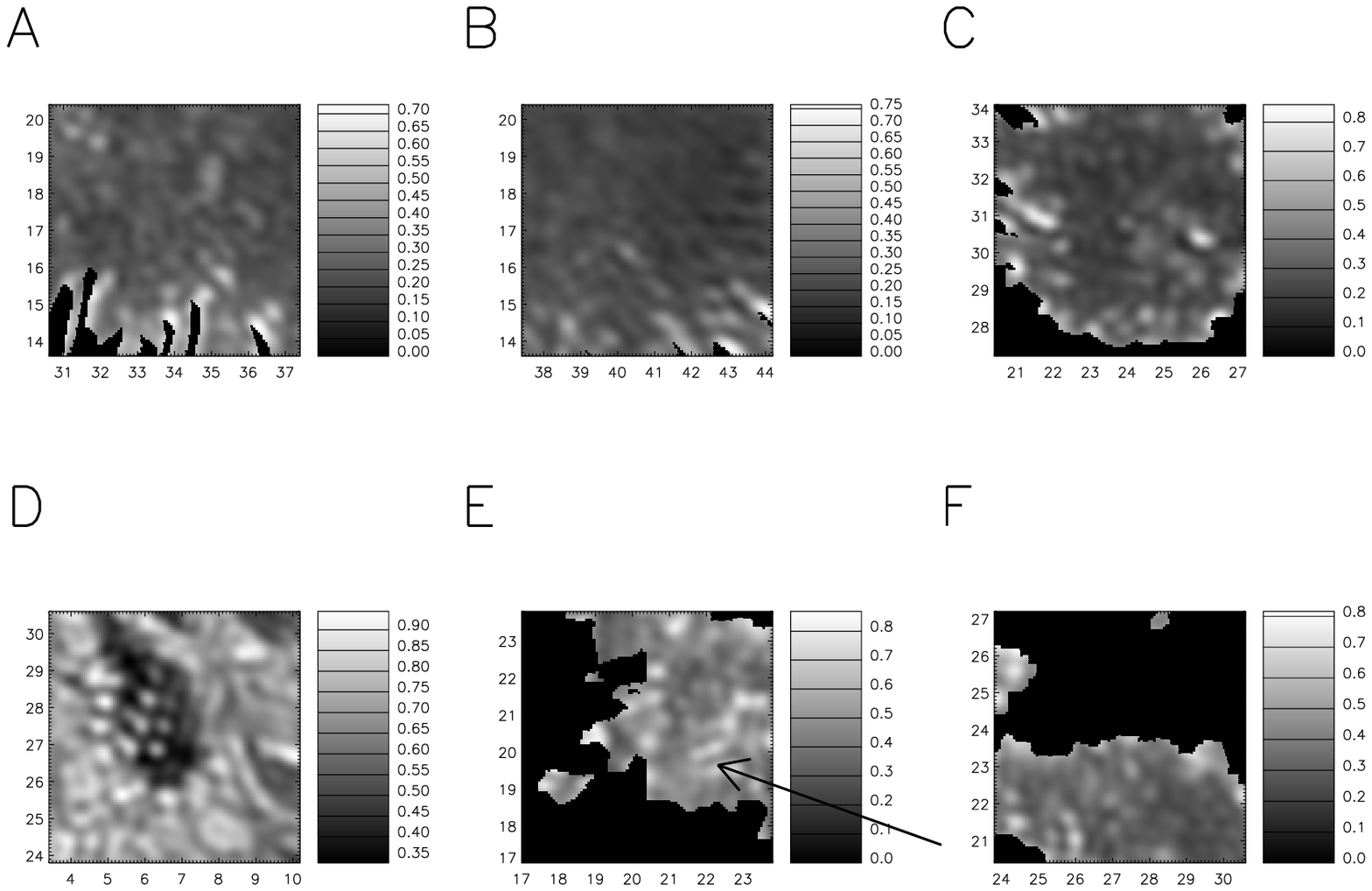} 
\caption{The zoomed in areas from Fig.\ref{pega}. Each panel corresponds to the square marked with the same letter. In the panels A, B,C, E and F we masked out the penumbral structures using the intensity of the image as the criteria. All pixels with the values larger than 0.7 of the maximum intensity value of the whole field of view are masked out. Panel D was not masked out because the pore does not contain the penumbra. Black arrow points at the structure presented in detail in Fig\ref{struktura}. The color bars represent the intensity of the panels all of them normalised at maximum intensity of the whole image from Fig\ref{pega}. 
}
\label{pegzum}
\end{figure}

\begin{figure}
\includegraphics[scale=0.9]{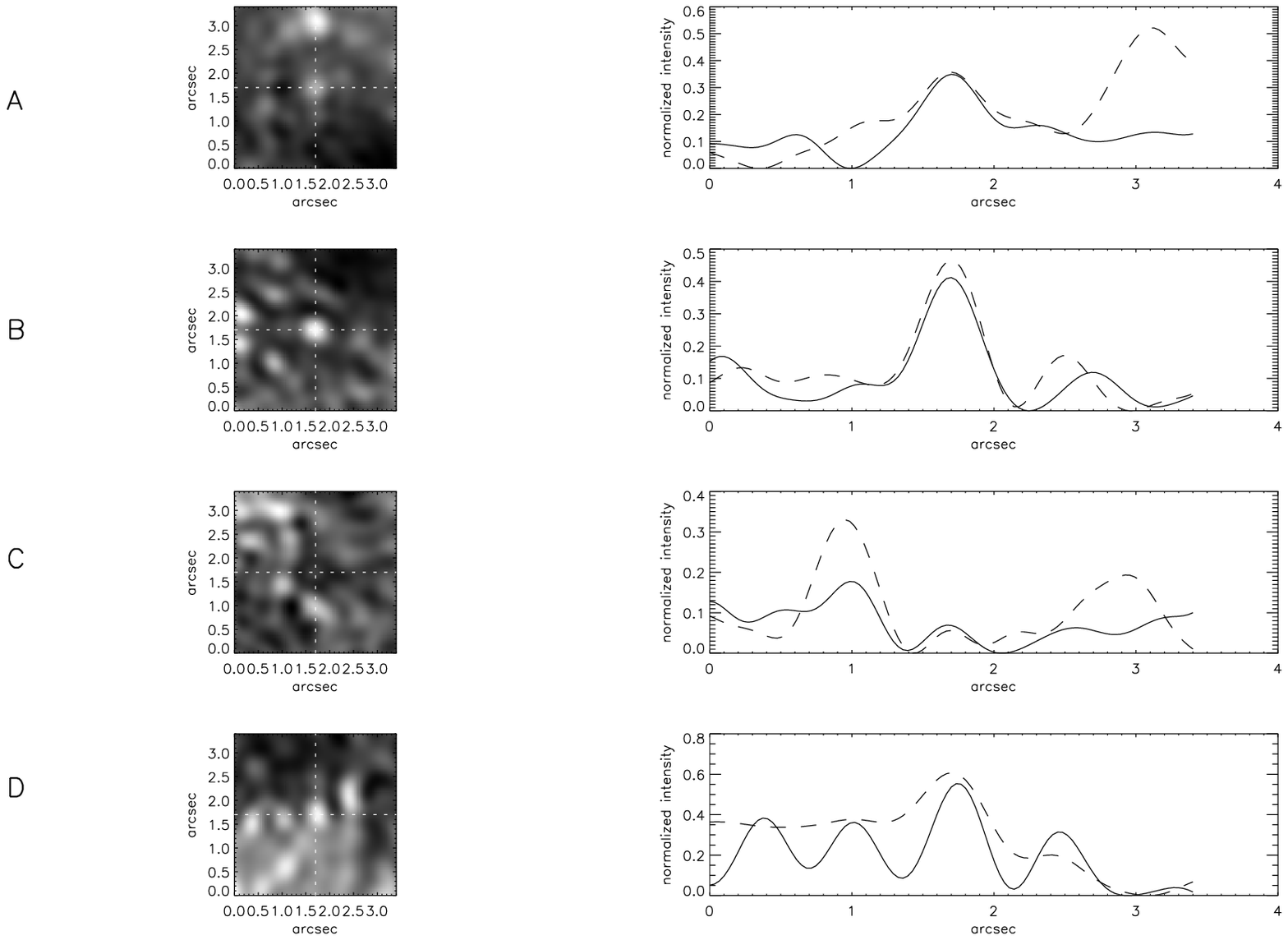} 
\caption{Zoomed UDs with the corresponding intensity profiles. Panel A represent UD located in Fig.\ref{pegzum}A. The panel consist of the intensity  map of a UD, with the UD in the middle of the frame, and a graph of two intensity profiles from the frame. Profiles are obtained from locations marked with doted lines in a frame. Solid line represents profile across X axis, while dashed line represent profile across Y axis. Profiles are made  pronounced by subtracting the minimum from the profile and then normalising the profile on the maximum intensity value from the frame itself. Panel B represents UD located in Fig.\ref{pegzum}B. Panels  C and D represent UDs  located in Fig.\ref{pegzum}A.}
\label{strukture}
\end{figure}

In six zoomed in areas (Fig.\ref{pegzum}) UD's appeared to have mostly circular shape. This is hard to notice in low contrast area of A i B areas, unless one inspects light profiles. The background intensities around UDs vary, posing an additional problem in resolving a dilemma;  is the observed structure a single UD with the dark lane or several UDs close together.  Different levels of umbral background intensity make hard to accurately measure FWHM of the structure and to distinguish one structure from another. NST resolution cannot answer the following question: "Are we seeing one UD surrounded with ones of smaller intensity or we see one bigger structure that has different intensity levels across its surface?" (Fig.\ref{strukture}A). Of course, there are UD that are clearly defined as separate objects, circular, and  without any visible substructure (Fig.\ref{strukture}B). In the parts of the umbra that are brighter, the NST can even detect very faint UDs with the qualities of a single structure (Fig.\ref{strukture}C). 

\begin{figure}
\includegraphics[scale=0.7]{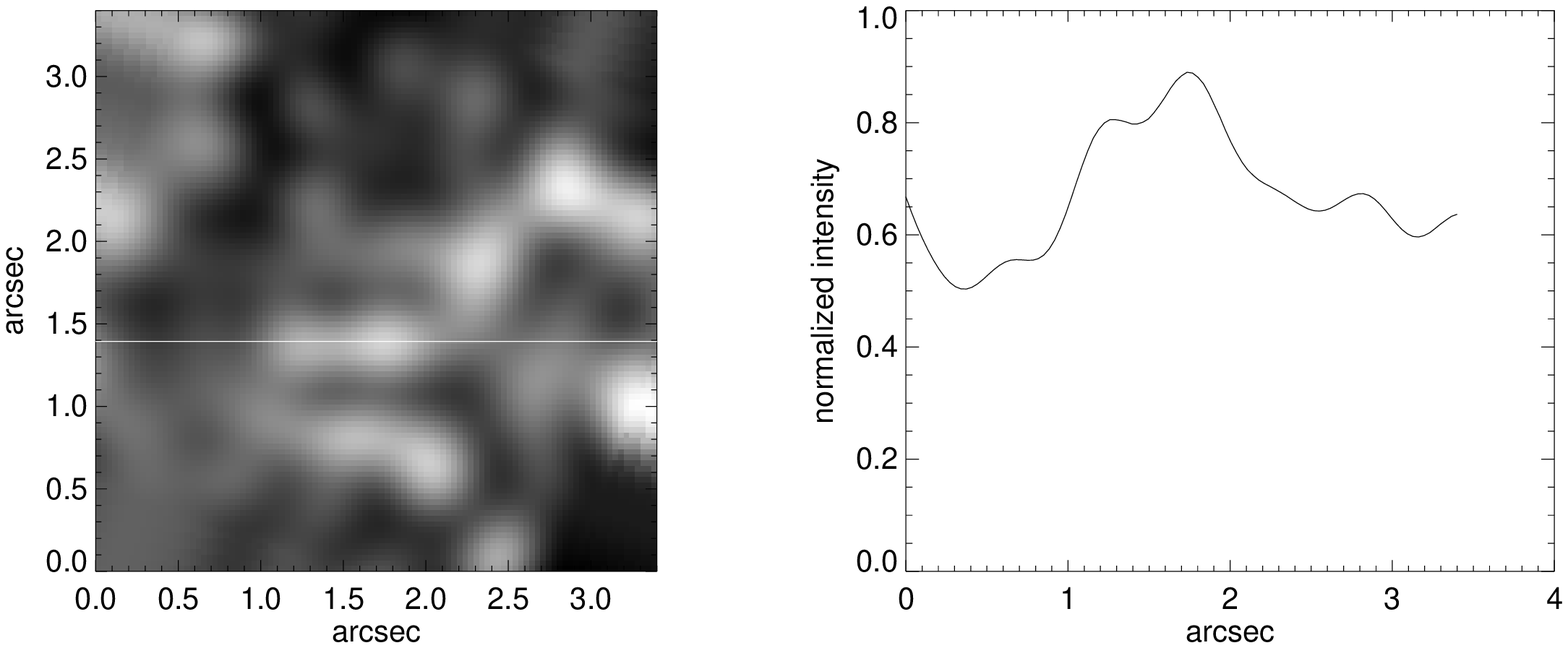} 
\caption{The structure with visible dark lane from  Fig.\ref{pegzum}E. Left panel presents the image of the structure. Image was rotated so that part of the structure is horizontal, white horizontal line represents the location of the intensity profile in the image. Right panel represents the intensity profile across the horizontal part of the structure.
}
\label{struktura}
\end{figure}

In the high contrast areas, (Fig.\ref{pegzum}D,E and F) circular shapes seems to be more apparent. However, Fig.\ref{pegzum}E and F show some elongated structures. The two structures at Fig.\ref{pegzum}E located at coordinates $\sim (22.5'', 19.5'')$ are the most similar to the assumed shape of the model (Sch\"{u}ssler and V\"{o}gler , 2006).  The shape smaller in size and similar in shape and the intensity distribution was discussed in Rimmelle (2008, Fig.4). However, the twin structure from our dataset is most likely a collection of several UDs located close to each other and poorly resolved with NST resolution in TiO spectral line. This line of thoughts occurs when one inspect intensity profiles of  those structures.  Fig.\ref{struktura} present the same elongated structure visible in Fig.\ref{pegzum}E. Intensity profile from Fig.\ref{struktura} shows that the part of the structure has two intensity peaks located very close to each other. Such profile might indicate two possibilities: first, a single structure with the dip in the intensity in the middle;  second,  two close individual structures that cannot be resolved with achieved angular resolution. 

\section{Discussion}

The UDs observed in TiO 705.68 nm spectral line had median diameter of $\sim 0.5''$, larger than result from model by  Bharti et al. (2010). This result is also larger than results of Watanabe et al. (2009, 2010) who reported the UDs size of $\sim 0.3"$ using Fe I 630 nm and Fe I 709.04 nm spectral lines respectively; and from the results of Sobtka et al. (1999) and Hamedivafa (2009) who reported size of $0.3"$ using broad band filters centred at spectral lines of 525.7 nm and 542.5 nm. However, the result is  in agreement with result of Rimmele (2008) who reported that size of the UDs varies from $0.2"$ to $"0.5"$ in G-band line. The difference in size might come from use of the TiO spectral line. This line has lower contrast than G-band resulting in the shallower light profiles across intensity map. Thus, measuring the size of the structure using FWHM might lead to larger results (Fig.\ref{strukture}), additional problem was establishing of the background intensity and thus isolating a single structure.  Medium intensity difference between the darkest pixel in the background of the UD and the brightest pixel in UD was maximum 44\% (in the pores). \par
Determination of the background intensity is problematic. Intensity levels differ  along the umbra. At this point it is not clear is this caused by topology of the umbra itself, or plethora of smaller structures in the umbra that are not fully resolved. One of the possible way to see into this problem is use of the 2D spectrometry with the high angular resolution instrument (such as are at the moment: NST at BBSO, USA; GREGOR at Tenerife observatory, EU; space based missions: HINODE and SUNRISE and eventually ATST at Hawaiian observatory, USA). Such data would provide informations along several heights revealing the shape of the umbral substructures in 3 spatial dimensions. \par

Significant number of UDs existed longer than our time series, limiting the measured median existence time to 35 minutes, but in the larger umbra, $60$\% of the UDs had shorter existance time.  This result is in agreement with the work of Sobotka et al. (1999), but in disagreement with works of Hamedivafa (2008) and Watanabe et al. (2009). \par

Although UDs observed in this dataset were larger than the UDs predicted by model, no dark lane was observed. There are two possible explanations: 
\begin{itemize}
\item TiO 705.68 nm spectral line is formed at the optical height where the dark lane do not appear. Sch\"{u}ssler and V\"{o}gler (2006) state that dark lane is starting to appear at the top of a plume that forms an UD. There is a possibility that TiO spectral line covers the optical heights lower than the ones covered  Iron spectral lines used in previous works. However, at this moment there is no research that would establish correct difference between formation height of the G-band and TiO bands. 

\item Second possibility is that resolution achieved in this work in this spectral line is not enough to resolve dark lanes. Rimmele (2008) made estimate of the size using different spectral lines. The TiO spectral line is weaker and thus produces less contrast (Sinha \& Tripathi, 1991a, 1991b). There is a possibility that in this spectral line those substructures appear smaller than reached diffraction limit due to the lower contrast of the line itself, and thus there were removed by applied filtering. 
\end{itemize}

The solution for this problem will be reached with the observations in G-band line using the NST 1.6 meter telescope. At this moment a conclusion can be made that  in TiO spectral line at 705.68 nm, with the diffraction limit of $0.1''$ umbral dots substructures cannot be observed. Moreover, although the TiO spectral line is seen as good line to study the umbra, experience with this data set obtained with the average atmospheric seeing condition indicate that best course of the action would be use of the stronger spectral lines that are capable to produce better contrast in photometry. \par

\it{Thanks are due to the anonymous referee whose comments helped improving this work. I wish to thank N. Gorceix for extensive help with the understanding an optical systems, and J.T. Villepique for help with the English. Also I wish to thank J. Varsik and R. Fear for the help with the observations.}

\bibliographystyle{abbrv}
\bibliography{simple}

\begin{thebibliography}{}

\bibitem[Andic et al. 2011]{ja11}Andi\'c, A.; Chae, J.; Goode, P. R.; Cao, W.; Ahn, K.; Yurchyshyn, V.; Abramenko, V., 2011, ApJ, 731, 29

\bibitem[Bharti et al. 2010]{bharti10}Bharti, L., Beeck, B., Sch\"{u}ssler, M., 2010, A\&A, 510, 12

\bibitem[Berger et al. 2004]{berger04}Berger, T.E., Rouppe van der Voort, L.H.M., L\"{o}fdahl, M.G., Carlsson, M., Fossum, A., Hansteen, V.H., Marthinussen, E., Title, A., Scharmer, G., 2004, A\&A, 428, 613

\bibitem[Cao et al. 2010]{cao10}Cao, W., Gorceix, N., Coulter, R., Woeger, F.,  Ahn, K., Shumko, S., Varsik, J., Coulter, A., Goode, P.R., 2010,  Proc. SPIE, 7735, 77355V-77355V-7
\bibitem[Chae and Sakurai 2008]{jongchul08}Chae, J., Sakurai, T. 2008, ApJ, 689, 593

\bibitem[Cheung et al. 2010]{cheung10}Cheung, M.C.M., Rempel, M., Title, A.M., Sch\"{u}ssler, M., 2010, ApJ, 720, 233

\bibitem[Hamedivafa 2008]{hame08}Hamedivafa, H., 2008, SoPh, 250, 17

\bibitem[Heinemann et al. 2007]{heinemann07}Heinemann, T., Nordlund, \AA., Scharmer, G.B., Spruit, H.C., 2007, ApJ, 669, 1390

\bibitem[Heinemann et al. 2006]{heinemann06}Heinemann, T., Dobler, W., Nordlund, \AA., Brandenburg, A., 2006, A\&A, 448, 731

\bibitem[Ortiz et al. 2010]{ortiz10}Ortiz, A., Bellot Rubio, L.R., Rouppe van der Voort, L., 2010, ApJ, 713, 1282

\bibitem[Rempel et al. 2009]{rempel09}Rempel, M., Sch\"{u}ssler, M., Cameron, R.H., Kn\"{o}lker, M., 2009, Sci, 325, 171

\bibitem[Riethm\"{u}ller et al. 2008]{Riethmuller08}Riethm\"{u}ller, T.L., Solanki, S.K., Lagg, A., 2008, ApJ, 678, L157 

\bibitem[Rimmele 2008]{rimmele08}Rimmele, T., 2008, ApJ, 672, 684

\bibitem[S\'{a}nches Almeida et al. 2004]{sanches04}S\'{a}nches Almeida, J., M\'{a}rquez, I., Bonet, J.A., Dom\'{i}ngues Cerde\~{n}a, I.F., Muller, R., 2004, ApJL, 609, 91

\bibitem[Sch\"{u}ssler and V\"{o}gler, 2006]{alex06} Sch\"{u}ssler, M., V\"{o}gler, A.,  2006, ApJ, 641, L73

\bibitem[Sinha and Tripathi, 1991]{sinha1991a}Sinha, K., Tripathi, B.M., 1991a, BASI, 19, 13

\bibitem[Sinha and Tripathi, 1991]{sinha1991b}Sinha, K., Tripathi, B.M., 1991b, BASI, 19, 23

\bibitem[Sobotka et al. 1999]{sobotka99}Sobotka, M., V\'{a}zquez, M., Bonet, J.A., Hanslmeier, A., Hirzberger, J., 1999, ApJ, 511, 436

\bibitem[von der L\"{u}he (1993)]{luhe93}von der L\"{u}he,O. 1993, A\&A, 268, 347

\bibitem[V\"{o}gler et al. 2005]{alex05}V\"{o}gler, A., Shelyag, S., Sch\"{u}ssler, M., Cattaneo, F., Emonet, T., Linde, T. 2005 A\&A, 429, 335

\bibitem[Watanabe et al. 2009]{watanabe09}Watanabe, H., Kitai, R., Ichimoto, K., 2009, ApJ, 702, 1048

\bibitem[Watanabe et al. 2010]{watanabe10}Watanabe, H., Tritschler, A., Kitai, R., Ichimoto, K., 2010, SoPh, 266, 5

\bibitem[W\"{o}ger et al. (2008)]{woeger}W\"{o}ger, F., von der L\"{u}he, O., Reardon, K. 2008, A\&A, 488, 375
\end{thebibliography}

\end{document}